\documentclass[twocolumn,aps,prl,letter,amsmath,amssymb,superscriptaddress]{revtex4}
\usepackage{bm,color,bbm}
\usepackage{amsmath}
\usepackage{hyperref,mathtools,graphicx,natbib}
\usepackage{bm,color,bbm}
\usepackage{hyperref, mathtools,graphicx,natbib}
\usepackage{graphicx}
\usepackage{amsfonts}    
\usepackage{graphicx}   
\usepackage{verbatim}   
\usepackage{color}      
\usepackage{subfigure}  
\usepackage{hyperref}   
\usepackage{natbib}
\usepackage{booktabs}
\usepackage{threeparttable}
\usepackage{dcolumn}
\usepackage{multirow}
\usepackage{array}
\usepackage{bm,color,bbm}
\usepackage{hyperref, mathtools,graphicx,natbib}
\usepackage{graphicx}
\usepackage{amsfonts}    
\usepackage{graphicx}   
\usepackage{verbatim}   
\usepackage{color}      
\usepackage{subfigure}  
\usepackage{hyperref}   
\usepackage{natbib}
\usepackage{booktabs}
\usepackage{threeparttable}
\usepackage{dcolumn}
\usepackage{multirow}
\usepackage{array}
\usepackage{ulem}

\newcommand{\beq}{\begin{equation}}
\newcommand{\eeq}{\end{equation}}
\newcommand{\bqa}{\begin{eqnarray}}
\newcommand{\eqa}{\end{eqnarray}}

\definecolor{maroon}{rgb}{0.7,0,0}

\definecolor{ngreen}{rgb}{0.3,0.7,0.3}

\definecolor{golden}{rgb}{0.8,0.6,0.1}

\graphicspath{ {./figures/} }

\begin{document}
\title{Synthesis and observation of optical skyrmionic structure in free space}
\author{Jie Zhu}
\affiliation{CAS Key Laboratory of Quantum Information, University of Science and Technology of China, Hefei, 230026, China}
\affiliation{CAS Center for Excellence in Quantum Information and Quantum Physics, University of Science and Technology of China, Hefei, 230026, China}

\author{Sheng Liu}
\affiliation{CAS Key Laboratory of Quantum Information, University of Science and Technology of China, Hefei, 230026, China}
\affiliation{CAS Center for Excellence in Quantum Information and Quantum Physics, University of Science and Technology of China, Hefei, 230026, China}

\author{Yong-Sheng Zhang}
\email{yshzhang@ustc.edu.cn}
\affiliation{CAS Key Laboratory of Quantum Information, University of Science and Technology of China, Hefei, 230026, China}
\affiliation{CAS Center for Excellence in Quantum Information and Quantum Physics, University of Science and Technology of China, Hefei, 230026, China}

\begin{abstract}
The skyrmion, which is characterised by a topological integer, is a structure that is topologically stable against local disturbances. The huge potential of skyrmions for use in magnetic storage systems has drawn considerable research interest among physicists. Recently, the optical skyrmion was discovered and has some excellent properties. However, these optical skyrmions have been observed, for example, in surface plasmons that consist of evanescent waves. This type of optical skyrmion is difficult to manipulate and also difficult to apply in practice. In this work, we realise several skyrmionic optical structures with different skyrmion numbers in a free-space linear optical system. Because of the convenience of operation using free-space optics, with the exception of the original applications of skyrmions, skyrmionic optical structures can also be applied widely, e.g. to enable manipulation of tiny objects or propagation over long distances.
\end{abstract}

\date{\today}
\maketitle


{\it Introduction---} Since it was first proposed by Skyrme in Ref. \cite{skyrme1962}, the concept of the skyrmion has been a subject of extensive study both theoretically and experimentally. Because of its topological excitation nature, the skyrmion is highly stable against local disturbances and is thus a promising candidate material for use in high-efficiency magnetic storage. Extensive research has been conducted on skyrmions \cite{wangkang2016} in condensed matter systems \cite{reviewskyrmion2020}, including quantum Hall systems \cite{jacek1996,shouchengzhang2011}, liquid crystals \cite{nagase2019} and Bose-Einstein condensates \cite{chunjiwang2010}.

Skyrmions can emerge through various mechanisms in real systems. For example, in magnetic systems \cite{kosuke2018}, there are long range magnetic dipolar interactions such as the Dzyaloshinskii-Moriya interaction \cite{beutier2017} that can produce skyrmions. The generated skyrmions have sizes ranging from several nanometres to several hundreds of nanometres. The skyrmion structure has also been generated via manipulation of optical fields. In Refs. \cite{tsesses2018,duluping2019}, evanescent electromagnetic fields were used to generate optical skyrmion lattices on the surface of a gold layer. The topological behaviour of the generated skyrmions was also demonstrated. 

To date, experimentally generated skyrmions have all been based on micro-interactions between light and matter or between magnetic spins. The topological character of these skyrmions is guaranteed by the underlying physical laws. Using the scheme presented in Ref. \cite{skyrmionbeams} as a basis, we have experimentally demonstrated a skyrmion-like structure in a linear optical system in free space \cite{skyrmionbeams}. For many of its potential applications, transformation or manipulation of the skyrmion is necessary, and thus production of skyrmions that can propagate freely in space has become an important task. We created a skyrmionic optical structure with six different skyrmion numbers. The corresponding skyrmion numbers that we extracted from the experimental results are consistent with the corresponding theoretical predictions. In addition, in our experiments, the skyrmion number is more accurate for larger distances. An important reason for this result is that the Laguerre-Gaussian (LG) mode approximation for the light emerging from the spatial-light modulator (SLM) is only valid for large distances.


{\it Theoretical framework---} Consider a magnetic spin model of a two-dimensional system. The unit magnetisation vector for the local spin is $\bm m(\bm r)=(m_1(\bm r),m_2(\bm r),m_3(\bm r))$, where $\bm r=(x,y)$. The skyrmion is a type of topological structure that is characterised by a topological number $N$, which is defined as the integral of the solid angle \cite{naoto2013,borge2020} as follows:
\begin{equation}
N=\frac{1}{4\pi}\int {\bm m\cdot(\frac{\partial {\bm m}}{\partial x}\times\frac{\partial {\bm m}}{\partial y})}d^2{\bm r}.
\label{skyrnumberdef}
\end{equation}
The skyrmion has a rotational symmetry about its centre and the unit magnetisation vector for the local spin can be written as $\bm m(\bm r)=(\cos\Phi(\phi)\sin\Theta(r),\sin\Phi(\phi)\sin\Theta(r),\cos\Theta(r))$, where we define $\bm r=(r\cos(\phi),r\sin(\phi))$ and $(\Phi(\phi), \Theta(r))$ represents the azimuth angle of the local spin. When the integral operation in Eq.~(\ref{skyrnumberdef}) is performed, we obtain $N=[\cos\Theta(r)]_{r=0}^{r=\infty}[\Phi(\phi)]_{\phi=0}^{\phi=2\pi}$. The boundary condition for a single skyrmion is that the spin points up at $r=0$ and points down at $r = \infty$ and that only one flip occurs between $r=0$ and $r=\infty$. The first term then gives $[\cos\Theta(r)]_{r=0}^{r=\infty}=2$ and the second term, which can be defined as the skyrmion vorticity, gives $m=[\Phi(\phi)]_{\phi=0}^{\phi=\infty}/2\pi$. This vorticity $m$ describes how many times the unit magnetisation vector covers the Bloch sphere as $r$ increases from zero to $r \to \infty$; therefore, we can define $m$ as the skyrmion number.

\begin{figure}[!t]   
    \centering
    \includegraphics[scale = 0.7]{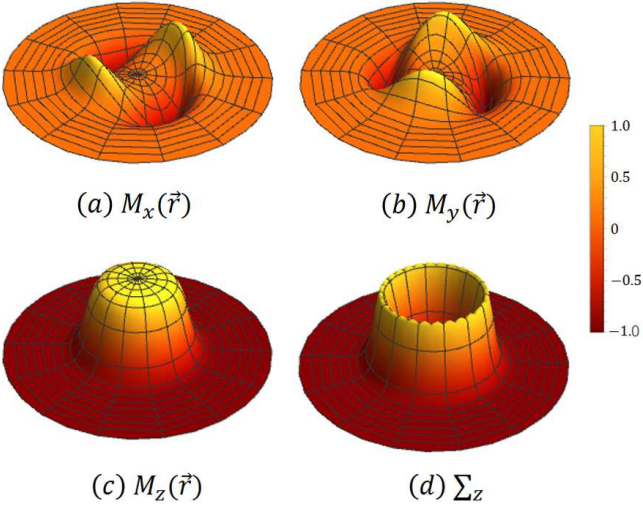}
    \caption{\textbf{Theoretical simulation.} The spatial distribution of the Poincare vector and the $z$-component skyrmion field is simulated based on Eq.~(\ref{initialstate}), Eq.~(\ref{skyrmionfield}) and Eq.~(\ref{poincare123}) with $l_1 = 0,~l_2 = 2$. $M_x(\bm r)$ \textbf{(a)} and $M_y(\bm r)$ \textbf{(b)} show rotational equality. If we select a closed path around the centre in the surface of $M_x(\bm r)$ or $M_y(\bm r)$, this path experiences a complete change from $-1$ to $1$ twice. The value of $M_z(\bm r)$ \textbf{(c)} is $1$ at the centre and $-1$ outside the bulge. From the centre outward to infinity along the radius, only one change from $1$ to $-1$ occurs. For the $\Sigma_z$, as shown in \textbf{(d)}, the value is $-1$ at the centre and $1$ within the closed ring. }
    \label{theoretical}
\end{figure}

Next, we consider the paraxial beam of light proposed in Ref.~\cite{skyrmionbeams}, where the beam state is (unnormalised)
\begin{equation}
|\widetilde{\Psi}(\bm r)\rangle=u_0(\bm r)|\varphi\rangle+u_1(\bm r){\mathrm{e}^{i\theta_0}}|\varphi^\perp \rangle ,
\label{initialstate}
\end{equation}
where $u_0(\bm r),u_1(\bm r)$ are both orthogonal LG modes \cite{allen1992} with orders of $l_1$ and $l_2 ~(l_2 > l_1)$ with $p=0$ respectively, and $\{|\varphi\rangle, |\varphi^\perp\rangle \}$ are both orthonormal polarisation states. After normalisation, the state vector becomes
\begin{equation}
|\Psi(\bm r)\rangle=\frac{|\varphi\rangle+{\mathrm e}^{i\theta_0}v(\bm r)|\varphi^\perp\rangle}{\sqrt{1+|v(\bm r)|^2}},
\label{normalizedstate}
\end{equation}
where $v(\bm r)=u_1(\bm r)/u_0(\bm r)$. 

As mentioned above, the skyrmionic field can be described using the effective magnetisation, which is the following local Poincare vector of light:
\begin{equation}
\bm M=\langle\Phi(\bm r)|{\bm \sigma}|\Phi(\bm r)\rangle,
\label{poincare}
\end{equation}
and $\bm \sigma=(\sigma_x,\sigma_y,\sigma_z)$ are the Pauli matrices in Cartesian coordinates. The $i$-th component of the skyrmion field can then be expressed as follows, based on the Poincare vector:
\begin{equation}
\begin{split}
\Sigma_i=\frac{1}{2}\epsilon_{ijk}\epsilon_{pqr}M_p(\bm r)\frac{\partial M_q(\bm r)}{\partial x_j}\frac{\partial M_r(\bm r)}{\partial x_k},
\label{skyrmionfield}
\end{split}
\end{equation}
where $\epsilon_{ijk}~(\epsilon_{pqr})$ is the total antisymmetric tensor. In our experiment, the propagation direction is along the $z$ direction and the outgoing light field is captured using a charge-coupled device (CCD) beam camera profiler. Therefore, we need only consider the $z$-th component of the skyrmionic field $\Sigma_z$ when calculating the associated skyrmion number:
\begin{equation}
N=\frac{1}{4\pi}\int \Sigma_z dxdy.
\label{skyrmionnumber}
\end{equation}
Here, in the case of the state shown in Eq.~(\ref{normalizedstate}), the skyrmion number is $N =\Delta l =  l_2 - l_1 > 0$.

To acquire the Poincare vector experimentally, we must actually measure the spatial distribution of the expected values of $\bm \sigma = (\sigma_x,\sigma_y,\sigma_z)$ as per Eq.~(\ref{poincare}). Therefore, we project the states into the six eigenstates of three Pauli operators. The corresponding measurement results are then recorded using the CCD profiler as $I_i(\bm r) $, where $i=\{I_{x1},I_{x2},I_{y1},I_{y2},I_{z1},I_{z2}\}$. The three components of the Poincare vector are then given by
\begin{equation}
M_i(\bm r) = \frac{I_{i1}(\bm r)-I_{i2}(\bm r)}{I_{i1}(\bm r)+I_{i2}(\bm r)},~i\in\{x,y,z\}.
\label{poincare123}
\end{equation}

Fig.~\ref{theoretical} shows a theoretical simulation of the spatial distribution of $M_i(\bm r)$ ($i = \{x,y,z\}$) and $\Sigma_z$, where $l_1 = 0$ and $l_2 = 2$, which indicates that the skyrmion number is 2. As Fig.~\ref{theoretical}(a) and Fig.~\ref{theoretical}(b) show, when traveling around the vortex once, one can experience two complete rotations of $\bm M$ between $[-1,1]$. The rotational equality of $M_x(\bm r)$ and $M_y(\bm r)$ is then obvious. The value of $M_z(\bm r)$ shown in Fig.~\ref{theoretical}(c) is $1$ at the centre and $-1$ outside the bulge. There is only one change, from $1$ to $-1$, from the centre to infinity along the radius. For $\Sigma_z$, the value in the centre is $-1$ and it then becomes $1$ in the closed ring. Therefore, when we calculate the skyrmion number based on the experimental data, the numerical integral is required to provide a truncation of the integral range. In the case where there are different values of $\Delta l $ with $l_2 > l_1$, the numbers of the petals in $M_x(\bm r)$ and $M_y(\bm r)$ are different and are equal to $\Delta l$, i.e. to the skyrmion number, while the shapes of $M_z(\bm r)$ and $\Sigma_z$ remain unchanged with different radii in the regions of variation.

\begin{figure}
\includegraphics[width=9cm]{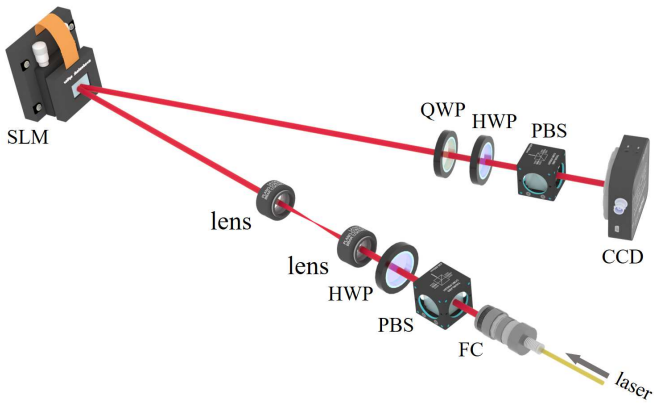}
\caption{\textbf{Experimental setup.} The continuous 780 nm laser beam is coupled into a single-mode fibre to introduce a pure Gaussian pump beam via the fibre coupler. The polarising beam splitter and the half-wave plate are used to prepare the initial polarisation state. Two lenses constitute the $4f$ system used to shape the transverse Gaussian profile. The orbital angular momentum degree of freedom is then obtained via the SLM. For the measurement part, the standard polarisation state analyser, which includes a half-wave plate, a quarter-wave plate and the polarising beam splitter, projects the polarisation state of the light into the eigenstates of the Pauli operators. Finally, the CCD profiler records the measurement results in the form of matrices. FC: fibre coupler; PBS: polarising beam splitter; HWP: half-wave plate; SLM: spatial light modulator; CCD: charge-coupled device beam camera profiler.}
\label{setup}
\end{figure}

{\it Experimental realization---} The experimental setup can be divided into two parts, comprising state preparation and measurement, as shown in Fig.~\ref{setup}. In the first part, a continuous laser beam with a centre wavelength of 780 nm is generated using a semiconductor laser; this beam is then coupled into a single-mode fibre via the fibre coupler (FC), which converts the divergent beam into a collimated beam, to obtain a pure Gaussian pump beam. Then, a polarising beam splitter (PBS) that only allows the horizontally polarised component to pass and a half-wave plate (H1 in Fig.~\ref{setup}) that can change the direction of the linear polarisation are used to prepare the initial polarisation state. Here, we fix the half-wave plate at $22.5^\circ$ such that the initial polarisation state after H1 is $|\psi_0\rangle=\frac{1}{\sqrt{2}}(|H\rangle +|V\rangle)$. Note that we select the basis $\{|H\rangle,|V\rangle\}$ as $\{ |\varphi\rangle,|\varphi^\perp\rangle\}$ here, where $|H\rangle~(|V\rangle)$ denotes the horizontal (vertical) polarisation. Next, the $4f$ system, which consists of two identical lenses with a focal length of 50 mm and a 40$\mu$m-diameter pinhole, reshapes the Gaussian beam’s transverse profile. To introduce the LG mode, the SLM is used to convert $|H\rangle|l=0\rangle$ into $|H\rangle|l_H\rangle$, where $l_H$ is a positive integer, and does not work for $|V\rangle$, i.e. $|V\rangle|l_V=0\rangle$ remains unchanged. We have now completed the state preparation and the state thus becomes $|\psi\rangle=\frac{1}{\sqrt{2}}(|H\rangle|l_H\rangle+|V\rangle|l_V=0\rangle)$. In our experiment, we select the values of $l_H$ from $\{2,4,6,8,10,12\}$. Therefore, in this case, $l_2 = l_H = \{2,4,6,8,10,12\}$ and $l_1 = l_V = 0$. Note here that, similar to the production of skyrmions in other systems, the SLM also realises spin-orbit coupling of the light beams.

The measurement part contains the standard polarisation state analyser and a CCD. The quarter-wave plate (QWP), the half-wave plate (HWP) and the PBS constitute the standard polarisation state analyser that is used for post-selection of the polarisation state. We adjust both the HWP and QWP to construct the post-selection states $\{|D\rangle, |A\rangle, |L\rangle, |R\rangle,|H\rangle, |V\rangle\}$, where $|D\rangle=\frac{1}{\sqrt{2}}(|H\rangle+|V\rangle)$, $|A\rangle=\frac{1}{\sqrt{2}}(|H\rangle-|V\rangle)$, $|L\rangle=\frac{1}{\sqrt{2}}(|H\rangle+i|V\rangle)$ and $|R\rangle=\frac{1}{\sqrt{2}}(|H\rangle-i|V\rangle)$, i.e. the eigenstates of the Pauli operators, and the measurement results correspond to $\{ I_{x1},I_{x2},I_{y1},I_{y2},I_{z1},I_{z2}\}$. The CCD then records the transverse profile of the beam that passes through PBS2 in the form of matrices \cite{scanfree,junliangjia2021,haijunwu2021,zhenxingliu2017}.

\begin{figure}
    \centering
    \includegraphics[scale = 0.5]{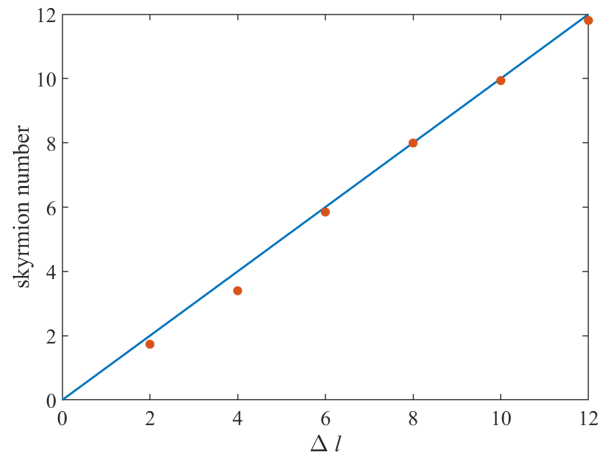}
    \caption{\textbf{Experimental data.} The skyrmion numbers with the different values of $\Delta l$ are calculated using Eq. (\ref{skyrmionnumber}). The error bars come from the resolution ratio of the CCD, that are too small to be visible in the figure. The blue line represents the theoretical prediction and the red dots where the errorbars are too small to present. The experimental values are $\{1.74\pm0.03, 3.40\pm0.04, 5.85\pm0.09, 7.99\pm0.06, 9.94\pm0.09, 11.81\pm0.03\}$ corresponding to $\Delta l = l_2-l_1 = \{2,4,6,8,10,12\}$, respectively.}
    \label{nskyrmiondelta}
\end{figure}

\begin{figure*}[!t]
    \includegraphics[width=18cm]{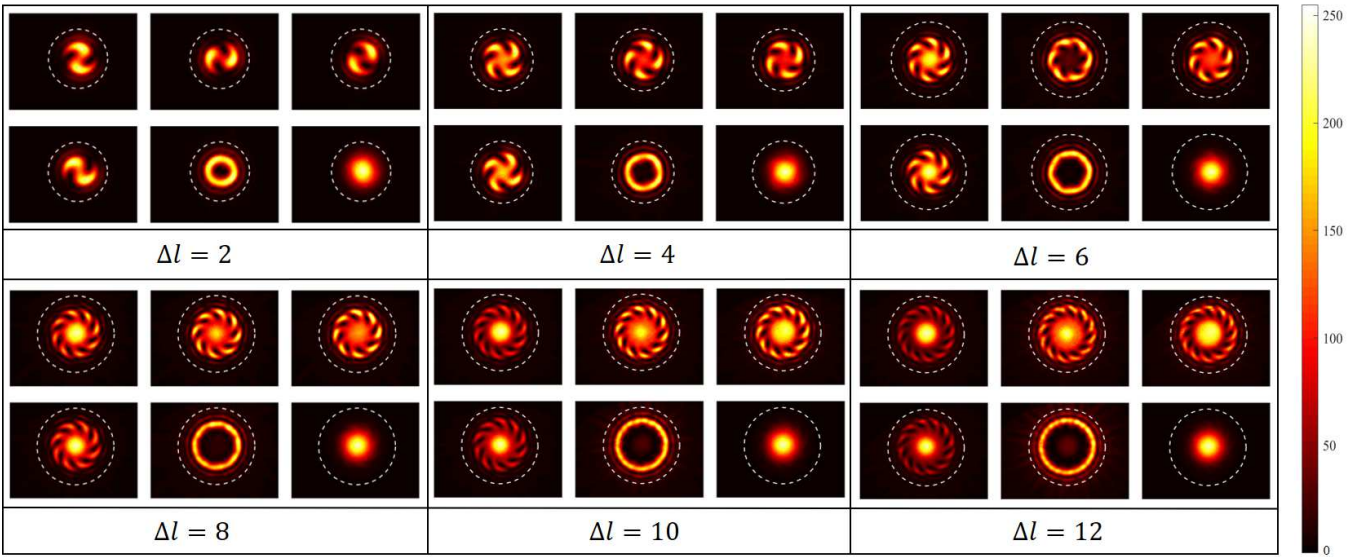}
    \caption{\textbf{Experimental data.} The measurement results are recorded using the CCD and correspond to $I_i(\bm r)$ with $i = \{x1,x2,y1,y2,z1,z2 \}$. We find that the $I_{z1}$ are the LG distributions with different orders $l$ and that the $I_{z2}$ are the same Gaussian distributions. The $\{I_{x1}, I_{x2}, I_{y1}, I_{y2}\}$ are then similar figures along different orientations. Here, $\Delta l  = |l_1-l_2| = l_1$ with $l_2=0$. The size of the skyrmionic structure then increases with increasing $l_1 $, i.e. there is a larger order between $l_1$ and $l_2$. The white dash line rings denote the integral ranges for each case when calculating the skyrmion number.}
    \label{photototal}
\end{figure*}

We compute the corresponding skyrmion numbers via Eq.~(\ref{skyrmionfield}) and Eq.~(\ref{skyrmionnumber}). As Fig.~\ref{nskyrmiondelta} shows, the skyrmion numbers for $\Delta l = l_2-l_1 = \{2,4,6,8,10,12\}$ are $\{1.74\pm0.03, 3.40\pm0.04, 5.85\pm0.09, 7.99\pm0.06, 9.94\pm0.09, 11.81\pm0.03\}$, respectively. The skyrmion numbers are not dependent on the intensity distribution; instead, they depend on the local ratio between the $|H\rangle$ component and the $|V\rangle$ component. Therefore, the skyrmion numbers will fluctuate severely when the intensity is too low in some areas. Most of the errors come from the LG mode generation process. In the experiments, we use a Gaussian beam to illuminate the SLM (liquid crystal on silicon (LCOS)-SLM, Hamamatsu, X10468), where the grey figure of the spiral phase is presented, to produce the LG mode. The beams after the SLM cannot actually be prepared accurately and in fact are hypergeometric Gaussian (HyGG) beams \cite{hyggmode}. The imperfections of the polarisation devices, including the wave plates and the PBS, introduce a slight distortion of the light spot. Additionally, the finite resolution of the CCD (Thorlabs, BC106N-VIS/M, $1360\times1024$ pixels) also limits the detection accuracy. The errorbars come from the finite resolution of the CCD, which are calculated via the local fluctuate sampling of each point.

The integral computations during the skyrmion number calculations are performed using Eq.~(\ref{skyrmionnumber}). Therefore, the integral boundary is reasonably important. As mentioned earlier, a Gaussian beam that is reflected by the SLM becomes HyGG beams that have more rings located outside the centre ring. When calculating the skyrmion number, we select the integral range from the centre ($r = 0$) to the places where the light density is small as the white dash line rings shown in the Fig.~\ref{photototal}. Then, we also calculate the skyrmion number using different integral ranges. It should be noticed that due to the imperfect of the devices, the centers of the experimental figure ($\{I_{x1}(\bm r),I_{x2}(\bm r),I_{y1}(\bm r),I_{y2}(\bm r),I_{z1}(\bm r)\}$) are not in the same point, where the tiny deviations are mainly caused by the waveplates in the measurement process, so that it is necessary to calibrate.

As shown in Fig.~\ref{photototal}, the detailed spatial distribution of $I_i(\bm r)$ has been recorded using the CCD. The patterns corresponding to $\{I_{x1}(\bm r),I_{x2}(\bm r),I_{y1}(\bm r),I_{y2}(\bm r),I_{z1}(\bm r)\}$ are shown to expand with increasing $\Delta l $. $I_{z2}(\bm r)$ is then always a Gaussian profile because $l_1 = l_V = 0$. The patterns of $\{I_{y1}(\bm r),I_{y2}(\bm r),I_{z1}(\bm r),I_{z2}(\bm r)\}$ are rotationally symmetrical.



{\it Discussion and conclusion---} A skyrmion in free space is different to a skyrmion in plasmonics in terms of its vector definition. Therefore, the authors of Ref.~\cite{tsesses2018} stated that there are no skyrmions in free space. This statement does not conflict with our results. The skyrmions in surface plasmonics use electric field vectors, but in the case presented here, we use the constructed vector of Pauli matrices $\bm \sigma$ (Poincare vector). The two orthogonal directions of $\{|H\rangle, |V\rangle\}~(\{|D\rangle,|A\rangle\},~\{|L\rangle,|R\rangle\})$ correspond to the antiparallel directions of the electric field vectors \cite{duluping2019}. 

The free-space skyrmionic optical structure actually offers several advantages over the plasmonic structure. First, the technology of free-space optics is mature \cite{robert2018,shengliu2018,jianchen2018,norrman2020}, which means that many adequate devices are available to control the skyrmions. In the experiments, we use the $4f$ system to reshape the Gaussian wavefront, the SLM to convert the wavefront, the wave plates to manipulate the polarisation states and the CCD to record images of the wavefront. Second, it is not difficult to obtain detailed information about the spatial distribution of the states via tomography \cite{tomographypra}. The state of each spin requires four projective measurements and the spins in the different positions can be measured simultaneously via tomography with the CCD. Therefore, to reconstruct the full spins in the optical field, only four projective measurements are required.


The free-space skyrmionic optical structure is a real and continuous distribution rather than discrete lattice points. In principle, there are no boundary limitations because the field becomes zero at the distances of the locations from the centre up to infinity. The continuous skyrmion-like structures can then be converted into discrete structures using the waveguide bundle \cite{jiho2020,adrian2019} and the optical fibre also represents a great tool to enable propagation of the skyrmions; this cannot be realised using plasmonic skyrmions. The interesting question of whether a skyrmion-like soliton \cite{stegeman1999} exists in a nonlinear medium arises naturally here. If the answer is yes, it will offer a more convenient and effective method to generate, transfer, store and manipulate skyrmions. On the other hand, the better quality of LG mode can improve the optical skyrmion structure.

In summary, we performed an experiment to realise a continuously-distributed bloch-type skyrmionic optical structure in free space. This structure is characterised by Poincare vectors, manipulated using linear optical elements and recorded using a CCD profiler. Skyrmionic optical structures in free space offer many promising advantages. These structures are convenient for propagation over long distances \cite{rozas1997,huancao2020,jianwang2012}. They can also be used as tools for information processing operations such as signal transformation and storage, according to the fundamental properties of skyrmions. Furthermore, the free-space optical skyrmion is also shown to be beneficial for coupling with other systems, including cold atoms and trapped ions. The free-space skyrmionic optical structure therefore provides a new way to manipulate other systems and may lead to different results similar to optical tweezers being used to control tiny objects and cold atom gases. For manipulation of cold atom gases in particular, such as Bose-Einstein condensates \cite{becsoc}, the free-space optical skyrmion may find many applications because of the polarisation dependence of atomic transitions and manipulations. This method of optical skyrmionic structure synthetization may be used to synthetize the material wave, such as the skyrmion in cold atom, and the skyrmion of electronic wave in free space.




{\it Acknowledgements---} We thank Kun Huang and Zhao-Di Liu for helpful discussions. This work was funded by the National Natural Science Foundation of China (Grants Nos.~11674306 and 92065113) and the Anhui Initiative in Quantum Information Technologies.

\hfill

\noindent {












\textbf{References}}



\begin{thebibliography}{42}

\bibitem{skyrme1962} T. H. R. Skyrme, {\it A unified field theory of mesons and baryons}, Nucl. Phys. \textbf{31}, 556 (1962).

\bibitem{wangkang2016} W Kang, Y Huang, X Zhang, Y Zhou and W Zhao, {\it Skyrmion-Electronics: An Overview and Outlook}, in Proceedings of the IEEE \textbf{104}, 2040 (2016).

\bibitem{reviewskyrmion2020} A. N. Bogdanov and C. Panagopoulos, {\it Physical foundations and basic properties of magnetic skyrmions}, Nat. Rev. Phys. \textbf{2}, 492 (2020).

\bibitem{jacek1996} J. Dziarmaga, {\it Statistics of skyrmions in quantum Hall systems}, Phys. Rev. B \textbf{53}, 12973 (1996).



\bibitem{shouchengzhang2011} X.-L. Qi and S.-C. Zhang, {\it Topological insulators and superconductors}, Rev. Mod. Phys. \textbf{83}, 1057 (2011).

\bibitem{nagase2019} T. Nagase, M. Komatsu, Y. G. So, T. Ishida, H. Yoshida, Y. Kawaguchi, Y. Tanaka, K. Saitoh, N. Ikarashi, M. Kuwahara, and M. Nagao, {\it Smectic Liquid-Crystalline Structure of Skyrmions in Chiral Magnet $\rm{Co}_{8.5}Zn_{7.5}Mn_4(110)$ Thin Film}, Phys. Rev. Lett. \textbf{123}, 137203 (2019).


\bibitem{chunjiwang2010} C. Wang, C. Gao, C.-M. Jian and H. Zhai, {\it Spin-Orbit Coupled Spinor Bose-Einstein Condensates}, Phys. Rev. Lett. \textbf{105}, 160403 (2010).



\bibitem{kosuke2018} K. Karube1, J. S. White, D. Morikawa, C. D. Dewhurst, R. Cubitt, A. Kikkawa, X. Yu, Y. Tokunaga, T. Arima,
H. M. Rønnow, Y. Tokura, Y. Taguchi, {\it Disordered skyrmion phase stabilized by magnetic frustration in a chiral magnet}, Sci. Adv. \textbf{4}, 7043 (2018).


\bibitem{beutier2017} G. Beutier, S. P. Collins, O. V. Dimitrova, V. E. Dmitrienko, M. I. Katsnelson, Y. O. Kvashnin, A. I. Lichtenstein, V. V. Mazurenko, A. G. A. Nisbet, E. N. Ovchinnikova, and D. Pincini, {\it Band Filling Control of the Dzyaloshinskii-Moriya Interaction in Weakly Ferromagnetic Insulators}, Phys. Rev. Lett. \textbf{119}, 167201 (2017).

\bibitem{tsesses2018} S. Tsesses, E. Ostrovsky, K. Cohen, B. Gjonaj, N. H. Lindner and G. Bartal, {\it Optical skyrmion lattice in evanescent electromagnetic fiels}, Science \textbf{361}, 993 (2018).

\bibitem{duluping2019} L. Du, A. Yang, A. V. Zayats and X. Yuan, {\it Deep-subwavelength features of photonic skyrmions in a confined electromagnetic field with orbital angular momentum}, Nat. Phys. \textbf{15}, 650 (2019).

\bibitem{skyrmionbeams}S. Gao, F. C. Speirits, F. Castellucci, S. F. Arnold, S. M. Barnett and J. B. G$\rm{\ddot{o}}$tte, {\it Paraxial skyrmionic beams}, Phys. Rev. A \textbf{102}, 053513 (2020). 

\bibitem{naoto2013}N. Nagaosa and Y. Tokura, {\it Topological properties and dynamics of magnetic skyrmions }, Nat. Nanotech. \textbf{8}, 723 (2013).

\bibitem{borge2020}B. G$\rm{\ddot{o}}$bel, I. Mertig and O. A. Tretiakov, {\it Beyond skyrmions: Review and perspectives of alternative magnetic quasiparticles}, Phys. Rep. \textbf{895}, 1 (2021).


\bibitem{allen1992}L. Allen, M. W. Beijersbergen, R. J. C. Spreeuw and J. P. Woerdman, {\it Orbital angular momentum of light and the transformation of Laguerre-Gaussian laser modes}, Phys. Rev. A \textbf{45}, 8185 (1992).


\bibitem{scanfree}Z Shi, M. Mirhosseini, J. Margiewicz, M. Malik, F. Rivera, Z Zhu and R. W. Boyd, {\it Scan-free direct measurement of an extremely high-dimensional photonic state}, Optica \textbf{2}, 388-392 (2015).

\bibitem{junliangjia2021} J. Jia, K. Zhang, G. Hu, M. Hu, T. Tong, Q. Mu, H. Gao, F. Li, C. Qiu, and P. Zhang, {\it Arbitrary cylindrical vector beam generation
enabled by polarization-selective Gouy phase
shifter}, arXiv: 2101.03738 [physics.optics].

\bibitem{haijunwu2021}H.-J. Wu, B.-S. Yu, Z.-H. Zhu, C. R. Guzm$\acute{a}$n, Z.-Y. Zhou, D.-S. Ding, W. Gao and B.-S. Shi, {\it Heralded generation of vectorially structured photons with high purity}, arXiv: 2101.06684 [Physics.optics].

\bibitem{zhenxingliu2017}Z. Liu, Y. Liu, Y. Ke, Y. Liu, W. Shu, H. Luo, and S. Wen, {\it Generation of arbitrary vector vortex beams on hybrid-order Poincar$\acute{e}$ sphere}, Photonics. Res. \textbf{5}, 15-21 (2017). 

\bibitem{hyggmode}A. Mawardi, S. Hild, A. Widera and D. Meschede, {\it ABCD-treatment of a propagating doughnut beam generated by a spiral phase plate}, Opt. Express \textbf{19}, 21205 (2011).

\bibitem{robert2018} R. C. Devlin, A. Ambrosio, N. A. Rubin, J. P. B. Mueller and R. Capasso, {\it Arbitrary spin-to–orbital angular momentum conversion of light}, Science \textbf{358}, 6365 (2018).

\bibitem{shengliu2018}S. Liu, S. Qi, Y. Zhang, P. Li, D. Wu, L. Han and J. Zhao, {\it Highly efficient generation of arbitrary vector beams with tunable polarization, phase, and amplitude}, Photonics. Res. \textbf{6} 4, 228 (2018).


\bibitem{jianchen2018}J. Chen, C.-H. Wan and Q.-W. Zhan, {\it Vectorial optical fields: recent advances and future prospects}, Sci. Bull. \textbf{63}, 54 (2018).


\bibitem{norrman2020}A. Norman, A. T. Friberg, and G. Leuchs, {\it Vector-light quantum complementarity and the degree of polarization}, Optica \textbf{7}, 93-97 (2020).

\bibitem{tomographypra} D. F. V. James, P. G. Kwiat, W. J. Munro, and A. G. White, {\it Measurement of qubits}, Phys. Rev. A \textbf{64}, 052312 (2001).

\bibitem{jiho2020} J. Noh, T. Schuster, T. Iadecola, S Huang, M. Wang, K. P. Chen, C. Chamon and M. C. Rechtsman, {\it Braiding photonic topological zero modes}, Nat. Phys. \textbf{16}, 989 (2020).

\bibitem{adrian2019} A. Hierro, M. M. Bajo, M. Ferraro, J. T. Arriola, N. L. Biavan, M. Hugues, J. M. Ulloa, M. Giudici, J. M. Chauveau, and P. Genevet, {\it Optical Phase Transition in Semiconductor Quantum Metamaterials}, Phys. Rev. Lett. \textbf{123}, 117401 (2019).



\bibitem{stegeman1999} G. I. Stegeman and M. Segev, {\it Optical Spatial Solitons and Their Interactions: Universality and Diversity}, Science \textbf{286}, 1518 (1999). 


\bibitem{rozas1997} D. Rozas, C. T. Law, and G. A. Swartzlander, {\it Propagation dynamics of optical vortices},  J. Opt. Soc. Am. B \textbf{14}, 3054-3065 (1997).

\bibitem{huancao2020} H. Cao, S.-C. Gao, C. Zhang, J. Wang, D.-Y. He, B.-H. Liu, Z.-W. Zhou, Y.-J. Chen, Z.-H. Li, S.-Y. Yu, J. Romero, Y.-F. Huang, C.-F. Li, and G.-C. Guo, {\it Distribution of high-dimensional orbital angular momentum entanglement over a $\rm 1~km$ few-mode fiber}, Optica \textbf{7}, 232 (2020).

\bibitem{jianwang2012} J. Wang, J.-Y. Yang, I. M. Fazal, N. Ahmed, Y. Yan, H. Huang, Y. Ren, Y. Yue, S. Dolinar, M. Tur and A. E. Willner, {\it Terabit free-space data transmission employing orbital angular momentum multiplexing}, Nat. Photonics \textbf{6}, 488-496 (2012).


\bibitem{becsoc} Y. J. Lin, K. J. Garc${\acute{i}}$a and I. B. Spielman, {\it Spin-orbit coupled Bose-Einstein condensates}, Nature (London) \textbf{471}, 83-86 (2011).

\end{thebibliography}
\end{document}